\documentclass[10pt,twocolumn,letterpaper]{article}

\usepackage{iccv}              

\definecolor{iccvblue}{rgb}{0.21,0.49,0.74}
\usepackage[pagebackref,breaklinks,colorlinks,allcolors=iccvblue]{hyperref}

\def\paperID{18} 
\def\confName{ICCV}
\def\confYear{2025}

\title{MSPT: A Lightweight Face Image Quality Assessment Method with Multi-stage Progressive Training}

\author{
Xiongwei Xiao\\
The Hong Kong Polytechnic University\\
{\tt\small xiongweixiaoxxw@gmail.com}
\and
Baoying Chen \thanks{Corresponding author}\\
Alibaba Group\\
{\tt\small 1900271059@email.szu.edu.cn}
\and
Jishen Zeng\\
Alibaba Group\\
{\tt\small jishen.zjs@alibaba-inc.com}\\
\and
Jianquan Yang\\
Shenzhen Campus of Sun Yat-Sen University\\
{\tt\small yangjq65@mail.sysu.edu.cn}
}

\begin{document}
\maketitle
\begin{abstract}
Accurately assessing the perceptual quality of face images is crucial, especially with the rapid progress in face restoration and generation. Traditional quality assessment methods often struggle with the unique characteristics of face images, limiting their generalizability. While learning-based approaches demonstrate superior performance due to their strong fitting capabilities, their high complexity typically incurs significant computational and storage costs, hindering practical deployment. To address this, we propose a lightweight face quality assessment network with Multi-Stage Progressive Training (MSPT). Our network employs a three-stage progressive training strategy that gradually introduces more diverse data samples and increases input image resolution. This novel approach enables lightweight networks to achieve high performance by effectively learning complex quality features while significantly mitigating catastrophic forgetting. Our MSPT achieved the second highest score on the VQualA 2025 face image quality assessment benchmark dataset, demonstrating that MSPT achieves comparable or better performance than state-of-the-art methods while maintaining efficient inference.
\end{abstract}    
\section{Introduction}
\label{sec:intro}
\begin{figure}
  \centering
    \includegraphics[width=\linewidth]{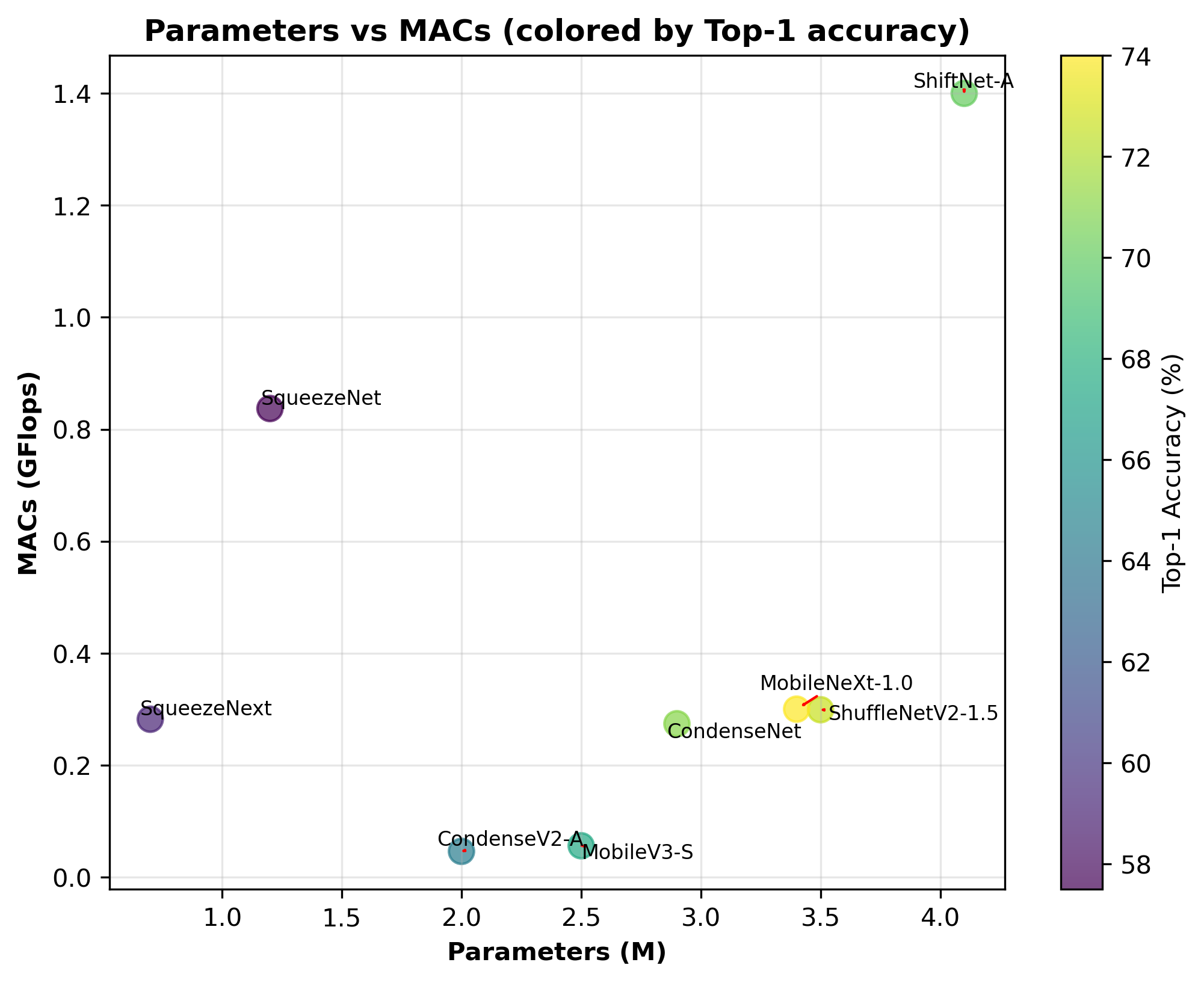}
    \caption{A comparison of various neural network architectures based on their computational complexity (MACs in GFlops) and model size (Parameters in M, no more than 5). Each point is colored by its classification Top-1 accuracy (\%).}
    \label{fig:paraandmacs}
\end{figure}
\begin{figure*}
  \centering
    \includegraphics[width=\linewidth]{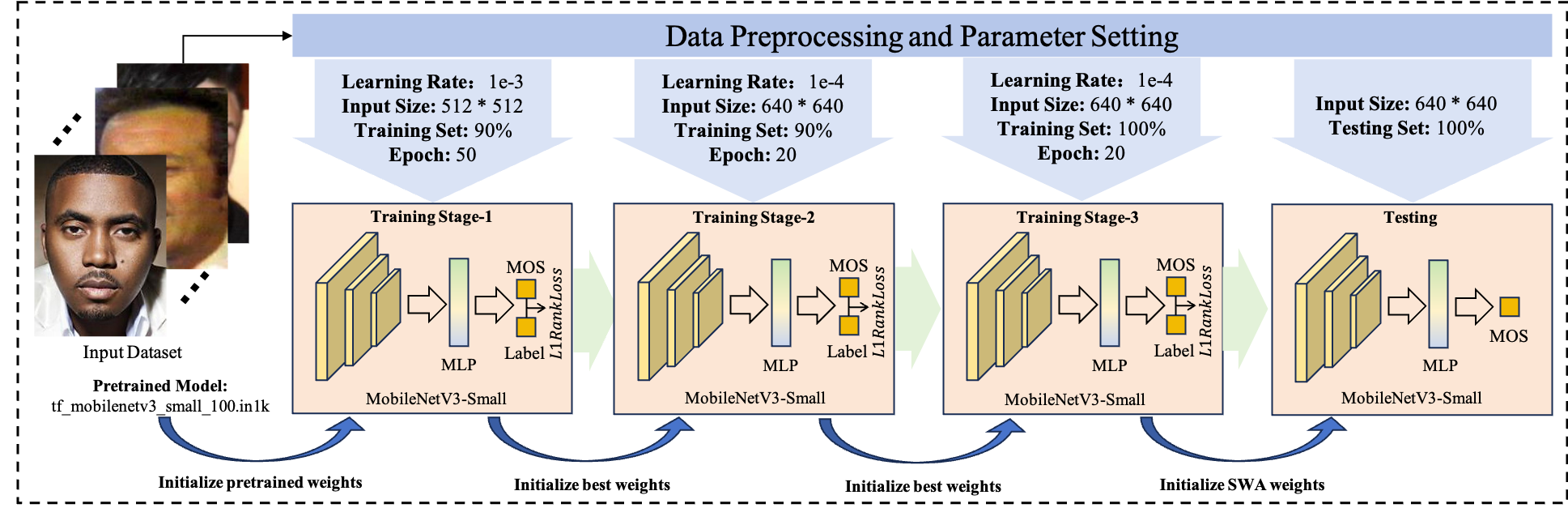}
    \caption{The pipeline of our method. The proposed multi-stage progressive training pipeline for face quality assessment. The model undergoes three training stages with progressively changing parameters (learning rate, input size, training set size) and is initialized with pretrained, best, and SWA weights respectively. A final testing phase evaluates the model's performance.}
    \label{fig:pipeline}
\end{figure*}

Assessing the perceptual quality of face images has become a crucial task in the era of digital media. It plays a fundamental role in improving face restoration algorithms, filtering low-quality training samples in generative models, and enabling robust performance in downstream applications such as face recognition and verification\cite{schlett2022face}. However, due to the intrinsic complexity of human faces, subjectivity of perceptual scores, and frequent occlusions or distortions in the wild, designing accurate and generalizable face image quality assessment (FIQA) models remains a challenging problem.

Traditional IQA methods, including full-reference and no-reference metrics, have shown limited effectiveness when directly applied to human faces\cite{schlett2022face}. These methods often fail to capture the semantic and structural subtleties that define perceptual face quality. In contrast, biometric face image quality assessment (BFIQA) focuses on recognition utility but may not reflect human-perceived quality degradation. To address this gap, recent research has shifted attention to general face image quality assessment (GFIQA), which aims to provide a more holistic and perceptually aligned quality measure for real-world face images.

In this work, we propose an efficient and scalable FIQA framework tailored for resource-constrained environments, where model size and computational cost are critical constraints. Our approach is motivated by the growing need for lightweight solutions that can be deployed on edge devices or mobile platforms without sacrificing performance. As shown in Figure \ref{fig:paraandmacs}, we show eight network models with less than 5M parameters, as well as their computational complexity and performance\cite{liu2024lightweight}. Specifically, balancing model compactness, computational efficiency, and performance, we adopted MobileNetV3-Small\cite{howard2019searching}. This architecture, featuring fewer than 5 million parameters and low computational complexity, is well-suited for real-time face quality assessment tasks.

To further enhance model performance and generalization, we introduce a three-stage progressive training strategy. We begin with low-resolution (512×512) inputs to allow rapid learning of global structures and coarse features, then gradually increase the input resolution to 640×640 in later stages to refine local perceptual cues. This curriculum-inspired progression helps the model adapt to the diverse image sizes in the dataset, which range from 200 to 1000 pixels in width. Additionally, we employ Stochastic Weight Averaging (SWA) to stabilize the final model and improve robustness across test scenarios. Our method aligns well with the goals of the VQualA workshop challenge, focusing on perceptual quality while satisfying strict efficiency requirements.

Through extensive experiments, we demonstrate that our progressive training pipeline leads to superior performance compared to standard fixed-resolution training, particularly in terms of both perceptual accuracy and inference efficiency. Our contributions show that well-designed training strategies, even when paired with lightweight backbones, can effectively bridge the gap between efficiency and quality in general FIQA tasks.

\section{Related work}
\label{sec:relatedwork}
\begin{figure}
  \centering
    \includegraphics[width=\linewidth]{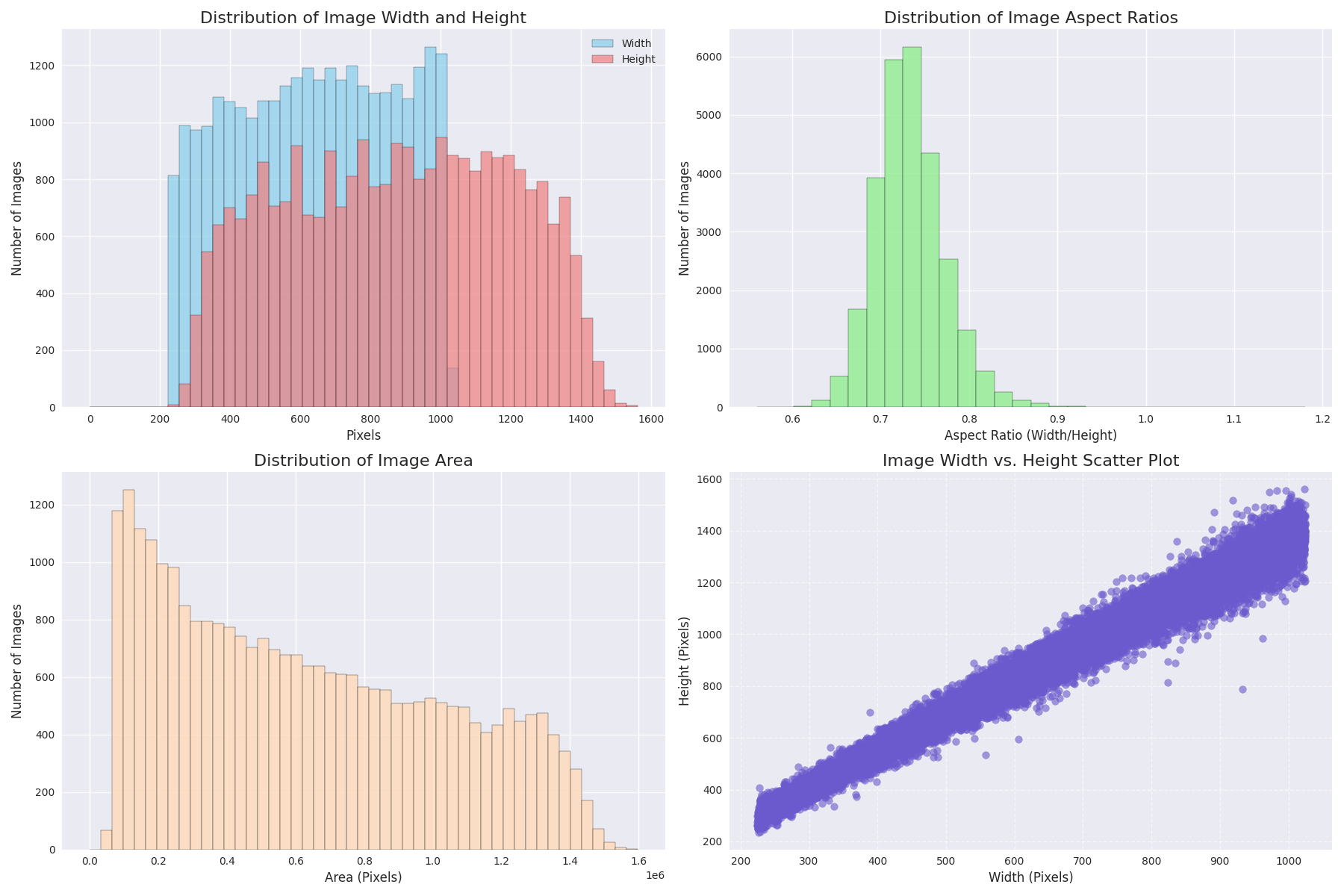}
    \caption{Statistical distributions of image properties within the dataset. (a) Distribution of image width and height. (b) Distribution of image aspect ratios (Width/Height). (c) Distribution of image area. (d) Scatter plot of image width vs. height, illustrating the correlation between dimensions}
    \label{fig:datasetstat}
\end{figure}
\subsection{Face Image Quality Assessment}
FIQA aims to automatically assign a quality score to a given face image, reflecting its utility in downstream face-related tasks such as face recognition\cite{fu2022deep} and AIGC-based face generation\cite{Porcile_2024_CVPR}. Since the performance of face recognition systems is highly dependent on input image quality, accurate FIQA is crucial for filtering low-quality samples and guiding system-level decisions. Unlike generic image quality assessment, FIQA focuses specifically on assessing the quality of single visible-light face images. Early FIQA approaches relied on handcrafted features or simple distortion-based metrics, but recent advances have shifted toward deep learning-based methods. Initial learning-based models, such as the work by \cite{best2018learning}, employed human annotations to define quality labels. Later research moved toward automatically generating quality anchors from recognition embeddings—for instance, \cite{hernandez2020biometric} used intra-class distances as quality scores, and \cite{xie2020inducing} leveraged paired face images to learn quality relationships. More recent methods, such as SDD-FIQA\cite{ou2021sdd} and CR-FIQA\cite{boutros2023cr}, adopt distribution-based or angular distance-based metrics to supervise training, enabling robust quality prediction without relying on subjective annotations. DSL-FIQA\cite{chen2024dsl} introduces a transformer-based framework with dual-set degradation representation learning and landmark-guided attention, supported by a comprehensive dataset CGFIQA-40k to address representation imbalances

\subsection{Lightweight Networks}
The pursuit of efficient deep learning has driven significant innovations in lightweight CNN design, where architectural refinements fundamentally redefine the computation-accuracy trade-off. Early breakthroughs emerged with MobileNets' introduction of inverted residual blocks and linear bottlenecks, which decoupled spatial and channel processing through depthwise-separable convolutions\cite{howard2017mobilenets,howard2018inverted,li2021searching}. This paradigm inspired subsequent variants: MobileNeXts\cite{zhou2020rethinking} refined the inverted structure via sandglass blocks to enhance gradient flow, while ShuffleNets\cite{ma2018shufflenet,zhang2018shufflenet} leveraged channel shuffling and CondenseNet adopted learned group convolutions to dynamically optimize feature reuse. Further efficiency gains arose from GhostNet's philosophy of generating "ghost" features through computationally cheap operations, effectively decoupling redundancy from representational capacity\cite{han2020ghostnet}. The culmination of these efforts materialized in EfficientNet's holistic compound scaling framework, which systematically coordinates depth, width, and resolution scaling to achieve Pareto-optimal performance\cite{tan2019efficientnet,tan2021efficientnetv2}. 

While deep learning has significantly advanced FIQA performance, the inherent computational and storage costs of complex deep models often hinder their deployment in resource-constrained environments, particularly for real-time applications. Concurrently, the extensive research into lightweight network architectures has demonstrated the feasibility of achieving competitive performance with greatly reduced computational footprints and parameter counts. Therefore, leveraging these advancements, there is a compelling need for lightweight Face Image Quality Assessment models. Such models would bridge the gap between high-accuracy assessment and practical deployability, enabling efficient and pervasive quality evaluation in various real-world face-related applications.

\section{Approach}
\label{sec:approach}

\begin{table*}[ht]
\centering
\small
\caption{Final Results of VQualA Face Quality Assessment using SRCC and PLCC in the Test phase} 
\label{tab:overallresults}
\begin{tabular}{llcccccc}
\toprule
\# & User & Score & SRCC & PLCC & FLOPS[G] & NumParams[M] & Runtime[S] \\
\midrule
1 & Team1 & 0.9664 (1) & 0.9692 & 0.9637 & 0.3313 & 1.1796  & 0.0097  \\
2 & \textbf{MSPT(Ours)} & 0.9624 (2) & 0.9624 & 0.9624 & 0.4687  & 1.5189  & 0.1920  \\
3 & Team3 & 0.9583 (3) & 0.9630 & 0.9535 & 0.4533  & 1.2224  & 0.0367  \\
4 & Team4 & 0.9566 (4) & 0.9600 & 0.9533 & 0.4985  & 2.0916  & 2.3660  \\
5 & Team5 & 0.9547 (5) & 0.9530 & 0.9564 & 0.4860 & 3.7171  & 0.0720  \\
6 & Team6 & 0.9432 (6) & 0.9431 & 0.9433 & 1.9198  & 3.2805  & 0.1920  \\
7 & Team7 & 0.9406 (7) & 0.9397 & 0.9415 & 0.4923  & 3.2805  & 0.1920  \\
8 & Team8 & 0.9334 (8) & 0.9413 & 0.9255 & 0.4097  & 3.2511  & 0.1920  \\
9 & Team9 & 0.9279 (9) & 0.9282 & 0.9275 & 0.4890  & 0.9513  & 0.1920  \\
10 & Team10 & 0.9242 (10) & 0.9262 & 0.9222 & 0.4895 & 4.0252  & 0.1920  \\
11 & Team11 & 0.9174 (11) & 0.9189 & 0.9159 & 0.4895  & 4.0252 & 0.1920  \\
12 & Team12 & 0.9038 (12) & 0.9118 & 0.8958 & 0.2235  & 4.7795  & 0.0500  \\
13 & Team13 & 0.8727 (13) & 0.8897 & 0.8557 & 0.5120  & 4.7242  & 0.1920  \\
14 & Team14 & 0.8432 (14) & 0.8529 & 0.8335 & 0.2852  & 1.3005 & 0.1500  \\
15 & Team15 & 0.8309 (15) & 0.8334 & 0.8283 & 0.3139  & 3.2511 & 0.1920  \\
16 & Team16 & 0.8052 (16) & 0.8109 & 0.7996 & 0.2635  & 1.0029  & 0.1920 \\
17 & Team17 & 0.6999 (17) & 0.7098 & 0.6900 & 0.8980  & 6.0523  & 0.1920  \\
18 & Team18 & -0.0355 (18) & -0.0321 & -0.0389 & 0.3140  & 3.2510 & 0.0040  \\
\bottomrule
\end{tabular}
\end{table*}

Face image quality assessment requires comprehensive understanding of facial features and structures, which necessitates adequate spatial resolution for accurate evaluation. However, training deep neural networks with high-resolution images from the beginning poses significant computational challenges and may lead to suboptimal convergence. To address this challenge, we propose a MSPT framework that gradually increases input resolution while progressively expanding the training dataset, enabling efficient training while maintaining high assessment accuracy, as shown in Figure \ref{fig:pipeline}.

\subsection{Progressive Resolution Learning Strategy}
We conducted an in-depth analysis of the VQualA face image dataset (see Figure \ref{fig:datasetstat} ). The analysis results reveal the multivariate distribution of image width, height, and aspect ratio, which indicates that the dataset contains face images of various sizes and ratios, with the ratio mainly concentrated around 0.7, and that a $640\times640$ image can cover the entire area of most images. Therefore, We reference the philosophy from FixRes\cite{touvron2019fixing}, which employs higher resolutions after conventional training for fine-tuning the classifier or final layers. For face image quality assessment, complete facial information is crucial for accurate evaluation. Therefore, we adopt a progressive training pipeline that starts with $512\times512$ resolution and subsequently increases to $640\times640$ resolution. This approach enhances performance while saving training time and computational resources.
The progressive resolution strategy can be formulated as:
\begin{equation}
    \mathcal{R} = \{r_1, r_2\} = \{512, 640\}
\end{equation}
where $r_1 < r_2$ represents the resolution sequence used in different training stages.

\subsection{Three-Stage Progressive Training Framework}

Our MSPT framework consists of three carefully designed stages, each serving a specific purpose in the progressive training process:

\textbf{Stage 1: Foundation Training}
In the first stage, we train the model using 90\% of the training dataset at $512\times512$ resolution for 50 epochs with learning rate 0.001. This stage establishes foundational feature representations:

\begin{equation}
\theta_1 = \arg\min_{\theta} \frac{1}{|D_1|} \sum_{(x_i, y_i) \in D_1} L(f_\theta(T_{512}(x_i)), y_i)
\end{equation}
where $D_1$ represents 90\% of the training set, $T_{512}$ denotes the random crop transformation to $512\times512$ resolution, $L$ is the loss function. $x_i$ and $y_i$ denote input image and label respectively.

\textbf{Stage 2: Resolution Enhancement}
The second stage increases the input resolution to $640\times640$ and fine-tunes the model for 20 epochs with reduced learning rate 0.0001:

\begin{equation}
\theta_2 = \arg\min_{\theta} \frac{1}{|D_1|} \sum_{(x_i, y_i) \in D_1} L(f_\theta(T_{640}(x_i)), y_i)
\end{equation}
where $\theta$ is initialized with $\theta_1$ from Stage 1, $D_1$ represents the same dateset as Stage 1.

\textbf{Stage 3: Full Dataset Fine-tuning}
The final stage utilizes the complete training dataset at $640\times640$ resolution for another 20 epochs with learning rate 0.0001:
\begin{equation}
\theta_3 = \arg\min_{\theta} \frac{1}{|D_{full}|} \sum_{(x_i, y_i) \in D_{full}} L(f_\theta(T_{640}(x_i)), y_i)
\end{equation}
where $\theta$ is initialized with $\theta_2$ from Stage 2, where $D_{full}$ represents the entire training dataset.

\subsection{Continual Learning Strategy}

Drawing from the principles of \cite{zhang2022continual}, our MSPT framework incorporates a continual learning strategy that addresses catastrophic forgetting while enhancing generalization capability. The progressive data augmentation and learning rate scheduling naturally implement continual learning principles.

\textbf{Progressive Data Introduction}: We deliberately use 90\% of training data in Stages 1 and 2 to first establish robust foundational quality assessment capabilities through focused learning (90\% data), then introduce the complete dataset in Stage 3. This strategy prevents catastrophic forgetting by gradually exposing the model to new information:

\begin{equation}
\mathcal{D}_{stage} = \begin{cases}
0.9 \cdot \mathcal{D}_{full} & \text{if stage} = 1, 2 \\
\mathcal{D}_{full} & \text{if stage} = 3
\end{cases}
\end{equation}

\textbf{Adaptive Learning Rate Scheduling}: Our learning rate strategy follows a high-to-low progression that aligns with continual learning principles. The conservative learning rates in later stages preserve previously learned representations while adapting to new data patterns:

\begin{equation}
\eta_{stage} = \begin{cases}
0.001 & \text{if stage} = 1 \\
0.0001 & \text{if stage} = 2, 3
\end{cases}
\end{equation}

\textbf{Knowledge Transfer Mechanism}: Inspired by the paradigm in \cite{ma2021remember}, all weights from previous stages serve as initialization for subsequent stages:

\begin{equation}
\theta_{stage+1}^{(0)} = \theta_{stage}^{(final)}
\end{equation}

This creates a knowledge transfer mechanism that maintains previously learned knowledge while adapting to new training conditions.

\subsection{Stochastic Weight Averaging for Model Ensembling}

To further improve generalization capability and assessment effectiveness, we apply Stochastic Weight Averaging (SWA) to the model weights from the third stage:

\begin{equation}
\theta_{SWA} = \frac{1}{K} \sum_{k=1}^{K} \theta_k
\end{equation}
where $\theta_k$ represents the model weights at the $k$-th checkpoint during Stage 3. This averaging process reduces variance and improves the robustness of the final model.

\begin{table*}[ht]
\centering
\small
\caption{Component ablation study results evaluating the impact of different training strategies and optimization techniques on the performance of our Face Quality Assessment model in the \textbf{development} phase.}
\label{tab:component_ablation}
\begin{tabular}{llcccc}
\toprule
\ Training Strategy & SWA & Full Fine-tuning & SROCC & PLCC & Score \\
\midrule
Two-stage (Direct $640\times640$) &  &  & 0.9827 & 0.9901 & 0.9864 \\
Two-stage (Direct $640\times640$) & \checkmark & & 0.9829 & 0.9901 & 0.9865 \\
Three-stage (512→640)  &  & \checkmark & 0.9826 & 0.9902 & 0.9864 \\
Three-stage (512→640)  & \checkmark & \checkmark & 0.9829 & 0.9902 & \textbf{0.9866} \\
\hline
\end{tabular}
\end{table*}

\begin{table*}[ht]
\centering
\small
\caption{Comparison of two-stage and three-stage progressive training strategies in the final \textbf{test} phase}
\label{tab:progressive_training}
\begin{tabular}{llcccc}
\toprule
Training Strategy & SROCC & PLCC & Score & Improvement \\
\midrule
Two-stage (Direct $640\times640$) & 0.9604 & 0.9618 & 0.9611 & - \\
Three-stage (512→640) & \textbf{0.9624} & \textbf{0.9624} & \textbf{0.9624} & \textbf{+0.0013} \\
\hline
\end{tabular}
\end{table*}

\subsection{Training Configuration and Optimization}

\textbf{Model Architecture}: We utilize the lightweight MobileNetV3-Small model (``tf\_mobilenetv3\_small\_100.in1k'') loaded via the timm library, providing an excellent balance between computational efficiency and assessment accuracy.

\textbf{Loss Function}: We adopt L1RankLoss\cite{wen2021strong} as our primary loss function, specifically designed for ranking-based quality assessment:

\begin{equation}
L = L_{mae} + \lambda \cdot L_{rank}.
\label{eq:overall}
\end{equation}
where $L_{mae}$ and $L_{rank}$ represent mean absolute error loss and pair-wise ranking loss, respectively, $ \lambda $ is a parameter that balances the two losses. The $L_{rank}$ is computed between all pairs $(i,j)$ as follows:

\begin{equation}
\mathcal{L}_{\text{rank}} = \frac{1}{n^2}\sum_{i=1}^{n}\sum_{j=1}^{n}\max\left(0, |y_i - y_j| - e(y_i,y_j) \cdot (\hat{y}_i - \hat{y}_j)\right)
\end{equation}
where the pairwise sign function $e(y_i,y_j)$ ensures correct ranking direction:
\begin{equation}
e(y_i,y_j) = 
\begin{cases} 
1, & y_i \geq y_j \\
-1, & \text{otherwise}
\end{cases}
\end{equation}
where $y_i$ and $\hat{y}_i$ denote the ground truth and predicted quality scores for image $i$, respectively.

\textbf{Optimization Strategy}: We use AdamW optimizer with CosineAnnealingLR scheduler (period=5) and batch size 256. Training is conducted on 4 NVIDIA 1080Ti GPUs using AMP mixed precision training.

\textbf{Data Augmentation}: We apply \textit{HorizontalFlip}, \textit{RandomRotate90}, \textit{PadIfNeeded}, and \textit{RandomResizedCrop} to enhance model robustness.





\section{Experimental Evaluation}
\label{sec:experiments}
\subsection{Evaluation Metrics}
We follow the evaluation metrics adopted in VQualA, specifically using the \textbf{Final Score = $0.5 \times SRCC + 0.5 \times PLCC$} as our final optimization objective. The calculation formulas for SRCC and PLCC are as follows:

The Spearman's Rank Correlation Coefficient (SRCC) is computed by:
\begin{equation}
    \text{SRCC} = 1 - \frac{6 \sum_{i=1}^{n} d_i^2}{n(n^2 - 1)}
\end{equation}
where $d_i$ represents the difference between the ranks of the $i$-th pair of model-predicted and ground-truth scores, and $n$ is the total number of samples.

The Pearson's Linear Correlation Coefficient (PLCC) is calculated using:
\begin{equation}
    \text{PLCC} = \frac{\sum_{i=1}^{n} (X_i - \bar{X})(Y_i - \bar{Y})}{\sqrt{\sum_{i=1}^{n} (X_i - \bar{X})^2 \sum_{i=1}^{n} (Y_i - \bar{Y})^2}}
\end{equation}
where $X_i$ denotes the $i$-th model-predicted quality score, $Y_i$ represents the $i$-th ground-truth subjective quality score, $\bar{X}$ and $\bar{Y}$ are the respective means of the predicted and ground-truth scores, and $n$ is the total number of samples.
\subsection{Overall Performance}
Our model was primarily developed for participation in the VQualA 2025 Face Image Quality Assessment Challenge, which aims to evaluate perceptual quality of face images under in-the-wild conditions. This competition imposes dual requirements: optimizing perceptual quality assessment while maintaining strict efficiency constraints (no more than 5 million parameters and 0.5 GFLOPs computational cost) – critical factors for practical deployment of IQA methods. As demonstrated in Table \ref{tab:overallresults} comparing our model against competing entries, the proposed solution achieved the second-highest overall score in the challenge. This competitive result validates the effectiveness of our training methodology.

\subsection{Ablation Studies}
To validate the contribution of each ccomponent, we conduct detailed ablation studies on MobileNetV3-Small-100 in development and test phase, as shown in Table \ref{tab:component_ablation} and Table \ref{tab:progressive_training} respectively. The three-stage training strategy is consistent with the pipeline. The two-stage training strategy is described in detail as follows:
\begin{itemize}
\item Stage 1: 90\% training data, lr=1e-3, input size=512×512, fine-tuning for 50 epochs
\item Stage 2: Full training data, lr=1e-4, input size=640×640, fine-tuning for 20 epochs
\end{itemize}Ablation studies show that both SWA and the three-stage progressive training strategy can improve the overall score, demonstrating the effectiveness of our MSPT method.

\section{Conclusion}
\label{sec:conclusions}
We presented Multi-Stage Progressive Training (MSPT), a lightweight framework for face image quality assessment that effectively balances computational efficiency with assessment accuracy. Our three-stage progressive training framework combines resolution enhancement, dataset expansion, and continual learning principles to enable lightweight networks to achieve superior performance while mitigating catastrophic forgetting. Experimental results on the VQualA 2025 Challenge validate our approach's effectiveness. Using a compact MobileNetV3-Small architecture with fewer than 5 million parameters and under 0.5 GFLOPs, our method achieved the second-highest overall score, demonstrating comparable performance to state-of-the-art methods while maintaining strict efficiency constraints. The success of MSPT highlights the importance of training methodology in achieving optimal performance with lightweight architectures, making it particularly suitable for deployment on edge devices and mobile platforms.
{
    \small
    \bibliographystyle{unsrt}
    \bibliography{main}
}

\end{document}